 \documentstyle[12pt]{article}
\begin{document}
 \begin{center}
 {\bf Intrinsic phase-decoherence of electrons by two-level systems}\\
 \end{center}
 \begin{center}
 Vipin Srivastava$^\ast$ and Raishma Krishnan$^\dag$\\
 School of Physics, University of Hyderabad, Hyderabad 500046, India.
 \end{center}

 The fundamental problem of phase saturation of electrons in a
 disordered mesoscopic system at very low temperatures is
 addressed. The disorder in the medium has both static and
 dynamic components, the latter being in the form of two-level
 systems (TLSs), which becomes just about the only source of inelastic
 scattering in the limit $T\rightarrow 0$. We propose that besides
 the inelastic nature of scattering from the TLSs, the phase-shift
 of the electrons is also affected by the nature of tunneling in
 the TLSs. The tunneling becomes incoherent as $T$ decreases due
 to increasing long-range interactions among TLSs and affects the
 phase coherence of electrons scattering from them. The
 competition between this effect, which increases as $\sim
 T^{-1}$, and that of the scattering rate $\tau_{e-TLS}^{-1}$ behaving as
 $\sim T$ apparently governs the phase-shift of electrons.\\

 PACS number(s): 73.23-b, 72.15.Lh, 73.20Fz, 72.10-d.\\

 $\ast$ e mail: vpssp@uohyd.ernet.in \\

 $\dag$ e mail: vpssprs@uohyd.ernet.in
 \newpage
 \noindent1. \textbf{Introduction}\\

One of the fundamental properties of an electron under quantum
mechanical conditions is that the phase of its wavefunction
remains coherent over a length of time and space. The electron
loses the phase coherence if it interacts with its environment.
The coherence-decoherence transition defines the transition from
quantum mechanical behaviour of a `closed' system to the classical
behaviour of the same system when it is treated as `open'. What
should comprise `a system of interest' and its `environment' with
which it may be coupled, depends on how we want to define the
problem; the experiments are designed accordingly. For example in
the problem of movement of an electron in a random potential, the
electron and the impurities or the disordered potential comprise
the system of interest while other electrons, phonons and
two-level systems (TLSs) etc. form parts of the environment.

For the problem of an electron in a random potential it is now
established that the elastic scatterings of the electron under
study from the impurities do not dephase it, or randomize its
phase [1]. On each such scattering episode the momentum of the
electron changes and also the phase of the single-electron
wavefunction, but in a deterministic manner. That is the changes
are correlated and computable, in principle. Normally the phase
relaxation length, or the length over which the phase has
reversed, can be much larger than the elastic mean free path. The
inelastic scattering events due to electron-phonon (e-ph),
electron-electron (e-e) or e-TLS interactions, on the other hand
randomize the phase of the electron wavefunction. It was believed
until recently that in the limit of temperature $T\rightarrow 0$
the inelastic events could be minimized and consequently the
phase-decoherence time $\tau_{\phi}$ and the phase-coherence
length $L_{\phi}$ could be arbitrarily large.

Under normal conditions achievable in experiments a disordered
system can be viewed as a jigsaw of phase-coherent units, where
each unit is of mesoscopic length scale and contains many elastic
scatterers. To study the long-range phase-coherence one ought to
do experiments on a sample of mesoscopic size and weak
localization is an appropriate phenomenon to investigate for
analyzing different scattering processes for it is sensitive to
phase relaxation and also momentum relaxation. All electrons
incident in the same state acquire the same phase shift after
going through a given set of elastic scatterers so much so that a
time reversed course of the same collisions can restore the
original phase of the wavefunction. An inelastic collision in the
course destroys the phase memory of the electron irrevocably.\\

\noindent2. \textbf{Saturation of phase-decoherence time}\\

Some recent experiments [2,3] along these lines have shown that
$\tau_{\phi}$, and in turn $L_{\phi}$, approaches a
\textit{finite} temperature independent value below a temperature
that may lie between a few mK and 10K depending on the system
under study.  It has been observed in a wide variety of disordered
conductors that $\tau_\phi$ saturates to a value between $10^{-9}$
sec and $10^{-12}$ sec below $4K$ depending on the system. It is
further claimed that this phase decoherence is intrinsic in
character, that is, it should not depend on the coupling with
extrinsic environmental factors. This led to the suggestion that
the decoherence could be caused by the zero point oscillations of
the electrons [3]. This point of view was, however, refuted on the
simple ground that the energy of the zero point oscillations
cannot be transferred in the course of inelastic collisions of the
electron in question with the electrons around it.

Not withstanding the mechanism of dephasing, which is a matter of
intense debate, the result is undoubtedly of fundamental
importance and of far reaching consequences. A number of
phenomena, including electron localization, that depend on quantum
interference would require rethinking. As for localization, the
experimental result [3] poses a serious puzzle by indicating that
the zero temperature dephasing length $L_{\phi} (T=0)$ is much
smaller than the typical localization length $\xi$. This would
have the serious implication that large localized states (or weak
localization, WL) may not exist. There ought to be a competition
between quenched or static disorder and the factors that limit the
intrinsic decoherence time $\tau_{\phi}$ in deciding the extent of
localization even for $T\rightarrow 0$. Consequently there could
be a lower critical value of disorder below which localization may
not happen. (If $L_{\phi} \sim \frac {\xi}{(k_Fl)}$, localization
will occur only if $\xi < L_\phi$, i.e. $k_Fl<1$, i.e. the WL
regime for which $k_Fl \sim 1$, as it
is understood now, should actually be non-existent; $k_F$ is
the Fermi wave vector and $l$ is the elastic mean free path.)\\

\noindent3. \textbf{Sources of dephasing}\\

Phonons are the most obvious source of dephasing, but they have to
be ruled out in the temperature range in which $\tau_\phi$ has
been observed to saturate. The inelastic processes that may
persist at such low temperatures are e-e and e-TLS interactions.

While e-e interaction is generally considered to be a good
candidate as a phase randomizing agent at low temperatures it
should be noted that the samples in the experiments in question
have fairly low concentration of electrons, $\sim 10^{12} cm^{-2}$
[4]. The e-e interaction energy should be considerably low in
these samples as compared to that in earlier experiments when the
electron density used to be much higher. It is therefore not clear
as to what extent will these interactions be instrumental in the
dephasing phenomenon. These could at most be quasi-elastic which
would imply that there would have to be many collisions of this
kind before the electron's phase would change by $2\pi$.

We will investigate in some detail the interaction of electrons
with TLSs as a plausible mechanism of the phase decoherence. In
this scenario one studies the movement of an electron under the
combined influence of  static or quenched disorder and a dynamical
environment due to an atom or a group of atoms moving back and
forth between two locations in space which correspond to minimum
energy states seperated by a potential barrier in the
configuration space. While the movement of an electron under the
influence of a static disorder is diffusive, the dynamical
disturbances in space caused by TLSs can make the electron move
into another state inelastically even at $T=0$. We expect that the
inelastic scattering from TLSs should dominate at very low
temperatures when other sources of inelastic scattering have
diminished or become ineffective. Inelastic scattering from TLSs
as the source of decoherence of an electron has been discussed in
the literature [5-7], but we have a very different mechanism to
propose for the dephasing phenomenon.

Ovadyahu [5] first found in some of his indium oxide samples with
low resistivity $(k_Fl>2)$ that inelastic scattering time was much
shorter than the e-e interaction time. The results fitted well
when e-TLS scattering was invoked as the phase-breaking mechanism.
Zawadowski et al. [6] considered non-magnetic TLSs having
degenerate Kondo ground states. In a certain temperature range the
TLSs exhibit non-Fermi liquid (NFL) behaviour which apparently is
the signature of dephasing. They propose that the scattering of an
electron from a TLS of this type changes the state of the TLS and
thus the environment of the electron changes. This acts back on
the electron and causes the dephasing. Imry et al. [7] proposed a
perturbative theory in electron-environment interaction where
$\tau_\phi$ is much longer than the quasi-elastic scattering time
$\tau$. The scattered electron undergoes a phase-shift due to the
motion of the scatterer. The phase-shift changes the conductance
and also causes dephasing. If the defect motion is slower than the
time scale of electronic motion only the conductance changes, but
if the defect moves faster than the
electronic motion then dephasing also happens.\\

\noindent4. \textbf{Our proposal}\\

Our arguments are a bit along the lines of Imry et al. [7] but
only to start with. We propose a novel mechanism for the
saturation of $\tau_\phi$ which accounts for the $T$-independence
of $\tau_\phi$. First of all we propose that the phase-shift of an
inelastically scattering electron should depend not only on the
fact that the scattering it is undergoing is inelastic in nature
but also on the character of the scatterer. The latter should be
particularly significant in the case of dynamical scatterers like
TLSs, whose character depends on the nature of the tunneling --
whether it is coherent or incoherent.

We can formally write,
\begin{equation}
\tau_\phi = f(\tau_{in},c),
\end{equation}
where $c$ is a parameter signifying the character of the inelastic
scatterer. In the case of e-TLS scattering
$\tau_{in}(=\tau_{e-TLS})$ would predominently depend on
temperature and concentration of TLSs. While its dependence on
concentration of TLSs is obvious, we understand the dependence of
$\tau_{e-TLS}$ on $T$ like this: first of all note that the
tunneling will have to be assisted by phonons if $\lambda$ is
larger than a certain $\lambda_{min}$ [8,9] where $e^{-\lambda}$
represents overlap between wavefunctions in the two potential
wells of the TLS. Since $\lambda$ depends inversely on $T$, the
tunneling will be increasingly more difficult as $T\rightarrow 0$
and at some point it will become impossible. Below this value of
$T$ the system will freeze into one of the configurations
represented by the two wells. Thus as $T$ decreases the tunneling
rate of this type of TLSs decreases and consequently the
scattering rate, $\tau_{e-TLS}^{-1}$ decreases. In simple terms
the rate of tunneling becomes slower than the time scale of the
electron movement. However, if the TLS is such that $\lambda <
\lambda_{min}$, then tunneling continues to happen even if
$T\rightarrow 0$. Such TLSs are particularly significant for the
saturation of $\tau_\phi$.

We will now discuss the parameter $c$ in the eqn. (1). The
character of a TLS can depend considerably on its interaction with
other TLSs. Note that the motion of the tunneling entity (an atom
or bunch of them) produces a strain field. If the TLSs are far
away from each other or the temperature is not low enough to
preclude phonons, then the strain field may not affect other TLSs
because it may get weakened or dissipated by phonons. But if the
concentration of TLS is high or there are intermediary impurities,
then the strain field produced by one TLS can affect the nature of
oscillation of other TLSs if the temperature is so low that the
dissipation of the strain field by phonons can be ruled out. While
an isolated TLS generally oscillates between two wells coherently,
the interaction between TLSs can make the oscillations incoherent.
This crossover, which happens primarily as a function of
decreasing temperature will have its influence on the phase-shift
of the electron scattering off the TLS; it should add to the
change of phase of the electron, which is otherwise happening due
to its scattering from the TLSs.

If the mean interaction between the TLS is represented by an
energy $J$ then one can combine with it the local parameter
$\Delta_0$ ($\sim e^{-\lambda}$, representing the seperation
between the wells of a TLS on the configuration-axis) and describe
the TLS-TLS interaction by a dimensionless parameter [10]
\begin{equation}
\mu = J/\Delta_0.
\end{equation}
If $J$ dominates over the local coupling energy $\Delta_0$ (i.e
interaction between TLSs is as large as, or larger than, the
coupling between wavefunctions in the two wells of a TLS) then
$\mu$ will exceed $1$ and the local tunneling motion will become
incoherent. In glassy systems $\Delta_0$ will have a wide
distribution which can be $\propto k_BT$ [11]. Consequently $\mu$
can be treated as $\propto T^{-1}$. Note that $\tau_{e-TLS}$ is
also $\propto T^{-1}$ [12]. In the background of all the above
discussion, the implications of both $\mu$ and $\tau_{e-TLS}$
being $\propto T^{-1}$ are important in so far as the phase change
of the electron is concerned.

Let us consider the two situations, $\lambda > \lambda_{min}$ and
$\lambda < \lambda_{min}$, separately. Infact both the situations
are important at the same time because TLSs satisfying either
conditions would be present in the system. In the TLSs for which
$\lambda > \lambda_{min}$ since the tunneling must be assisted by
phonons, it will increasingly slow down with decreasing
temperature. As a result $\tau_{e-TLS}$ will increase at a rate
$\propto T^{-1}$. But, while the tunneling rate slows down the
{\textit{nature}} of tunneling changes from coherent to incoherent
at the same rate, namely $T^{-1}$ due to increasing TLS-TLS
interaction. If according to Stern et al. [13] e-TLS scattering is
turning from inelastic to quasi-elastic and then to elastic with
decreasing $T$ and is therefore able to change only the
conductance, the phase-shift in the scattered electron will be
mainly brought about by the latter factor discussed above, namely
the coherent-to-incoherent crossover in the nature of TLSs.

The TLSs with $\lambda < \lambda_{min}$, in which the tunneling
occurs in spite of the absence of phonons, change the phase of the
electron scattering off them even as $T\rightarrow 0$. This is
true independent of whether the energy exchanged in the scattering
process is sufficient for it to be fully inelastic. The phase of
the scattered electron in any case changes because the tunneling
in these TLSs also becomes incoherent with decreasing $T$ due to
long-ranged interactions among TLS.

The incoherence of the tunneling in both types of TLSs increases
with decreasing $T$ and accordingly the amount of the phase change
of the scattered electron increases. At sufficiently low $T$ when
the phase change is about $2\pi$ the decoherence time
$\tau_{\phi}$ will saturate and become temperature independent.\\

\noindent5. \textbf{Conclusions and comments}\\

We have argued that the phase decoherence time $\tau_\phi$ of
electrons in a disordered mesoscopic system saturates and becomes
$T$-independent at very low temperature due to the following two
complementing effects: on one hand with decreasing $T$ the e-TLS
scattering weakens and becomes less effective as an inelastic
process, but on the other hand this is compensated by the fact
that with decreasing $T$ the tunneling in TLSs becomes incoherent
due to increasing long-ranged interactions among TLSs and this
decoheres the electrons; both these effects balance each other at
the same rate, namely $T^{-1}$, which makes the saturated
$\tau_\phi$ independent of $T$. A good amount of further
experimentation is required to ascertain a number of finer points.

First of all it must be checked whether the systems that exhibit
$\tau_{\phi}$-saturation always have TLSs present in them. This is
necessary in order to establish that the suggested mechanism is
really a universal one. Experiments are required to identify the
types of TLSs, atleast the two types discussed above. Having done
this it is necessary to study in greater detail the
coherent-incoherent tunneling crossover as a function of $T$ and
whether both the above types of TLSs are affected equally as $T$
decreases. Experiments are also needed to ascertain if the e-e
interactions in {\it{dilute}} electron systems indeed involve
sufficient energy to be classified as inelastic at low
temperatures.

Finally it could well be, as suggested by Imry et al. [7], that
there may be just a temperature range over which $\tau_{\phi}$
remains saturated, and that at a $T << {\hbar/\tau_0}$ ($\tau_0$
being the saturation value of $\tau_{\phi}$) the $\tau_{\phi}$ may
diverge. Only very careful experimentation can resolve these
issues.

We may mention in passing that the rapid development of quantum
information processing [14] led to renewed interest in the study
of dephasing effects. The studies proposed above would be very
vital in this modern context for one of the major obstacles in the
way of implementation of quantum computers is the
relatively short dephasing time in the solid state devices.\\

 \textit{Acknowledgements:} RK is supported by the Council of
 Scientific and Industrial Research, New Delhi. VS is thankful to
 the Ministry of Information Technology, Government of India, for
 a grant.\\

 \noindent{\bf References}
 \begin{enumerate}
 \item R. Landauer, Phil. Mag. {\bf 21}, 863 (1970).

 \item S. Wind, M. J. Rooks, V. Chandrasekhar
 and D. E. Prober, Phys. Rev. Lett. {\bf 57},
 633 (1986); P. Mohanty and R. A. Webb, Phys. Rev. {\bf B55}, R13452 
(1997).

 \item P. Mohanty, E. M. Q. Jariwala and R. A. Webb,
 Phys. Rev. Lett. {\bf 78}, 3366 (1997); P. Mohanty and
 R. A. Webb, Phys. Rev. {\bf B55}, R13452 (1997).

 \item Y. U. Khavin, M. E. Gershenson, A. L. Bogdanov,
 Phys. Rev. Lett. {\bf 81}, 1066 (1998).

 \item Z. Ovadyahu, Phys. Rev. Lett. {\bf 52}, 569 (1984).

 \item A. Zawadowski, Jan Von Delft and D. C. Ralph, Phys.
 Rev. Lett. {\bf 83}, 2632 (1999).

 \item Y. Imry, H. Fukuyama and P. Schwab, Europhys.
 Lett. {\bf 47}, 608 (1999).

 \item P. W. Anderson, B. L. Halperin and C. M. Varma, Philos.
 Mag. {\bf 25}, 1 (1972); W. A. Phillips,
 J. Low Temp. Phys. {\bf 7}, 351 (1972).

 \item $\lambda_{min} \equiv ln(2\hbar\omega_0/\Delta E)$
 where $\hbar \omega_0$ is of the order of zero-point energy and
 $\Delta E$ is the difference between the minima of the two wells
 of the TLS, i.e. between the diagonal elements of the Hamiltonian
 for the TLS. The condition $\lambda > \lambda_{min}$ simply means
 that $\hbar \omega_0 e^{-\lambda}$, the off-diagonal element of
 the Hamiltonian, is $< \Delta E /2$. That is, $\Delta E$ is too
 large to allow any meaningful correction to the eigen values of
 the Hamiltonian by the off diagonal elements.

 \item A. Wurger, Z. Phys. {\bf B94}, 173 (1994);
 {\bf B98}, 561 (1995).

 \item C. Enss and S. Hunklinger, Phys. Rev. Lett. {\bf 79},
 2831 (1997).

 \item See e.g. R. Krishnan and V. Srivastava, Phys. Rev. {\bf B 59},
 R12747 (1999).

 \item A. Stern, Y. Aharonov and Y. Imry, Phys. Rev.
 {\bf A41}, 3436 (1990).

 \item D. P. DiVincenzo, in \textit{Mesoscopic electron transport},
 L. L. Sohn et al. (eds), Kluwer Academic Publishers, Dordrecht,
 1997; A Steane, Rep. Prog. Phys. {\bf 61}, 117 (1998).

 \end{enumerate}

 \end{document}